\documentclass[journal]{IEEEtran}
\ifCLASSINFOpdf
\else
\fi
\usepackage{amsfonts}
\usepackage{amsmath,amsfonts,amssymb,epsfig,subfigure,cite,url,multirow,booktabs}
\usepackage{afterpage,algorithm,algorithmic,mathrsfs}
\usepackage{amsthm,graphicx}

\usepackage{amsmath}
\begin{document}
%
\title{Energy Model for UAV Communications: Experimental Validation and Model Generalization}
%
%
%

\author{Ning Gao,~Yong Zeng,~Jian Wang,~Di Wu,~Chaoyue Zhang,~Qingheng Song,~Jiachen Qian, and~Shi Jin
\thanks{This work was supported in part by the Program for Innovative Talents and Entrepreneur in Jiangsu Province under Grant 1104000402, in part by the Research Fund by Nanjing Government under Grant 1104000396, in part by the open research fund of the National and Local Joint Engineering Laboratory of RF Integration and Micro-Assembly Technology under Grant Nos. KFJJ20180205, and in part by the NUPTSF Grant No. NY218113 \& No. NY219077. \textit{(Corresponding author: Yong Zeng.)}}
\thanks{N. Gao, Y. Zeng, D. Wu, J. Qian and S. Jin are with the National Mobile Communications
Research Laboratory, Southeast University, Nanjing 210096, China.}
\thanks{J. Wang and C. Zhang are with the College of Electronic and Optical Engineering \& College of Microelectronics, Nanjing University of Posts and Telecommunications, Nanjing 210023, China.}
\thanks{Q. Song is with the College of Electrical and Information Engineeing, Huaihua University, Huaihua 418008, China, and also with the Key Laboratory of Intelligent Control Technology for Wuling-Mountain Ecological Agriculture in Hunan Province, Huaihua 418008, China.}
}
\maketitle

\begin{abstract}
Wireless communication involving unmanned aerial vehicles (UAVs) is expected to play an important role in future wireless networks. However, different from conventional terrestrial communication systems, UAVs typically have rather limited onboard energy on one hand, and require additional flying energy consumption on the other hand, which renders energy-efficient UAV communication with smart energy expenditure of paramount importance. In this paper, via extensive flight experiments, we aim to firstly validate the recently derived theoretical energy model for rotary-wing UAVs, and then develop a general model for those complicated flight scenarios where rigorous theoretical model derivation is quite challenging, if not impossible. Specifically, we first investigate how UAV power consumption varies with its flying speed for the simplest straight-and-level flight. With about 12,000 valid power-speed data points collected, we first apply the model-based curve fitting to obtain the modelling parameters based on the theoretical closed-form energy model in the existing literature. In addition, in order to exclude the potential bias caused by the theoretical energy model, the obtained measurement data is also trained using a model-free deep neural network. It is found that the obtained curve from both methods can match quite well with the theoretical energy model. Next, we further extend the study to arbitrary 2-dimensional (2-D) flight, where, to our best knowledge, no rigorous theoretical derivation is available for the closed-form energy model as a function of its flying speed, direction, and acceleration. To fill the gap, we first propose a heuristic energy model for these more complicated cases, and then provide experimental validation based on the measurement results for circular level flight.
\end{abstract}


%
\IEEEpeerreviewmaketitle

\section{Introduction}
\IEEEPARstart{I}{ntegrating} unmanned aerial vehicles (UAVs) into terrestrial cellular networks is regarded as a win-win technology for both UAV industry and cellular operators \cite{8918497}. On one hand, UAVs with their respective missions can be connected to cellular networks as new aerial users for supporting their command and control as well as payload communications, a paradigm known as \textit{cellular-connected UAVs}. On the other hand, dedicated UAVs can be deployed as new aerial communication platforms such as flying base station and relays, to assist in terrestrial wireless communications, which is known as \textit{UAV-assisted wireless communication}.

Despite of the appealing benefits for integrating UAVs into cellular networks, there are still many practical challenges that hinder their large-scale usage. One critical issue is the limited onboard energy of UAV \cite{7888557}, which restricts the UAV endurance and hence operation time. Besides, different from that in conventional terrestrial communications, UAV energy consumption consists of two main parts: one is the conventional communication-related energy consumption due to, e.g., signal processing, circuits and power amplification; the other is the propulsion energy consumption to ensure that the UAV remains aloft and supports its movement. In general, the propulsion energy consumption depends on the flying status and is usually much more significant than communication-related power expenditure. Therefore, understanding how the UAV energy consumption varies with its flying status is of paramount importance for achieving energy-efficient UAV communications, i.e., maximizing the communication performance with a smart UAV energy expenditure.
\subsection{Energy Model Review}\label{sec:review}
The prior study on UAV energy model can be loosely classified into three categories. The first category simply applies the existing energy models of the ground vehicles or robots to UAVs \cite{7891919,6851894}. For example, in \cite{8254720}, by using the energy models of ground vehicles \cite{7891919} and terrestrial base stations \cite{7063976}, the placement of latency aware UAV base station was studied in heterogeneous networks with the energy capacity limitation. As DC motors are frequently adopted in ground robots to provide the body power, the energy model of a robot can be modeled as the sum of energy model of DC motor, which can be closely approximate by a six degree polynomial \cite{1302401}. While the power consumption of mobile robots moving on the ground are not applicable for UAVs due to their fundamentally different moving mechanisms. In particular, the movement of ground robots mainly needs to overcome the friction force, that of aerial vehicles needs to overcome the air-resistance force or drag force. The second category is the experimental-driven heuristic energy model based on measured/simulated data, \cite{7101619,7556979,8204378,7565126,8486942}. For example, in \cite{7101619}, the energy consumption of a specific drone is given as a function of its speed and operating conditions based on experiment result. Reference \cite{7848883} and \cite{8038869} have used the energy model in \cite{7101619} to study the energy-efficient UAV communications. Nevertheless, the energy model in \cite{7101619} is simply expressed as the integral of power over time, the lack of closed form expression limits its application. With the dynamic model of a lithium polymer battery, the relationship between energy consumption and quadrotor movement was experimentally studied in \cite{7556979,7565126,8486942}. For solar-powered UAV, the authors gave the solar cells energy consumption expression and verified it via the virtual flight evaluation system \cite{8204378}. The last category is the theoretical-driven energy model by kinematic and aircraft theory \cite{Helicopter,Helicopter2}, such as \cite{8102734,7991310,Phung2013Modeling,7925671,8264241}. For example, the authors in \cite{8102734} have defined an energy model as an integral function of the motor torque and rotor speed, where the motor torque is modeled as being proportional to the current through the torque constant. The steady state energy model assumes that the UAV is in steady state behavior most of the time \cite{7991310}. However, such model is limited on steady state flight, that is the external forces are in static equilibrium, yielding zero net acceleration. Furthermore, the force analysis is needed to obtain the steady state thrust and many complex parameters are required when using such model. The authors in \cite{Phung2013Modeling} modeled the energy consumption of a class of small convertible UAVs as a function of propeller's speed and torque, with six model parameters identified based on reported data. Moreover, the quadcopter power consumption estimation has been analyzed with the simulation of the quadcopter movement position vector \cite{7925671}. The propulsion energy model based on a integral function of propulsion and speed has been verified in \cite{8264241} by using the empirical parameters provided in \cite{BekmezciFlying}.

Note that most of the prior works either gave heuristic energy model without rigorous mathematical derivation, or used rather sophisticated models based on aerodynamic theory involving many implicit parameters that are difficult to interpret nor use in wireless communities. To address such issues, in our previous work \cite{7888557,8663615}, rigorous mathematical derivations have been performed to develop closed-form expressions for energy model with respect to the UAV flying status, such as speed and acceleration. Specifically, for a fixed-wing UAV, the energy model consists of two main components: the parasite power and the induced power. The parasite power, which increases in cubic with the flying speed, is required to overcome the parasite friction drag, and the induced power, which is inverse proportional to the flying speed, corresponds to the component required to overcome the induced drag. Different from fixed-wing UAVs, the energy consumption of rotary-wing UAVs contains an additional component, namely, blade profile power, which is needed to overcome the profile drag due to the rotation of blades. The energy model for rotary-wing UAV in straight-and-level flight with constant speed has been derived \cite{8663615}. Note that although such models are easy to interpret and have been extensively used for designing energy-efficient UAV communication systems, to our best knowledge, they have not yet been experimentally validated in practice. As a preliminary effort to fill this gap, in this paper, we have conducted extensive flight experiments to practically measure the UAV power consumption so as to validate the energy model of rotary-wing UAV developed in \cite{8663615}. Specifically, we first conducted the simplest straight-and-level flight experiment to validate the relationship between power consumption and flying speed for the developed theoretical energy model. As an effort for model generalization, we propose a heuristic energy model for the more complicated cases with arbitrary flight, and then provide experimental validations based on the measurement results of the circular level flight with uniform speed.
\subsection{Contributions}
In this paper, we validate the theoretical energy model of straight-and-level flight for rotary-wing UAVs via extensive flight experiments, then generalize the model to more complicated flight scenarios. The specific contributions of our work are summarized as follows:
\begin{itemize}
  \item For our recently developed theoretical energy model of rotary-wing UAVs in straight-and-level flight, we validate its correctness via extensive flight experiments. Specifically, with about 12,000 valid power-speed data points collected, model-based curve fitting is applied to find the key modelling parameters. In addition, to exclude the potential bias caused by the theoretical energy model, we further apply the model-free deep neural network to train energy consumption curve. The fitting curves form both methods match quite well with theoretical energy model, which provides the double validations about the model correctness.
  \item To generalize the energy model, we further extend the study to arbitrary 2-dimensional (2-D) flight. However, as there is no rigorous theoretical derivation available for the closed-form energy consumption as a function of its flying speed, direction, and acceleration, we propose a heuristic energy model for such more complicated scenarios. To validate the proposed heuristic energy model, we conducted flight experiment for the special circular level flight, where the measurement results gave good match with the generalized energy model.
  \item Some insights are obtained with the measurement results. For straight-and-level flight, the theoretical energy model better characterizes the energy consumption than the polynomial model. With the experiments performed, it is found that when the flight speed is under $14$ m/s, the energy consumption of the straight-and-level flight is about $350\thicksim 500$ W. For circular level flight, when the flight speed is low, the flight radiuses have insignificant impact on power consumption, and with the flight speed increases, the impact of flight radius on power consumption becomes more significant.
\end{itemize}

Note that some prior efforts on UAV, some flight experiments have also been pursued. In \cite{7101619}, the flight time and energy in a random area was verified by actually flying the area with a quadcoper. Three experiments including hover, steady-state ascend/descend and cyclic straight line were conducted to verify the energy model in \cite{7991310}. In \cite{7565126}, the authors tested the Parrot AR. drone's power and energy consumption for moving horizontally, vertically upwards and downwards by connecting the battery contacts to a DC power supply. Testing more flight behaviors, a series of experiments focused on understanding the impact of several factors on energy consumption of UAVs have been conducted in \cite{8486942}. Considering the payloads, the authors in \cite{4650856} tested the endurance estimation model with variations in payload mass and payload power consumption by using a custom-developed quadrotor and autonomous ceiling attachment system. The impact of ground power consumption, communication with the ground station, movement, speed and payload has been analyzed with the data based power consumption model \cite{8690666}. In \cite{7556979}, the fixed time and fixed distance experiments were performed in a closed environment to avoid windy conditions. However, most of such existing experiments aim to obtain an energy model expression by fitting measurement data, rather than verifying a theoretically derived energy model by measurement data, i.e., \cite{7565126,8486942,8690666}. Furthermore, limited work has considered the flight acceleration/decleration phase in the experiments, which typically has a high impact on model validation \cite{7991310}. To fill this gap, the main objective of our current work is to validate the theoretical model derived in our previous work \cite{8663615} via the flight experiments, based on which a more general model for sophisticated flight scenarios is developed.

\section{Theoretical Energy Model Overview and Flight Experiment Setup}\label{sec:1}
In this section, we first provide a brief overview on the theoretical energy model derived in \cite{8663615} and give some insights. Then, we design three flight experiments and discuss the details of the corresponding data processing.
\subsection{Energy Model for Straight-and-Level Flight}
\begin{figure}[!ht]
 \centering
  \includegraphics[width=8cm]{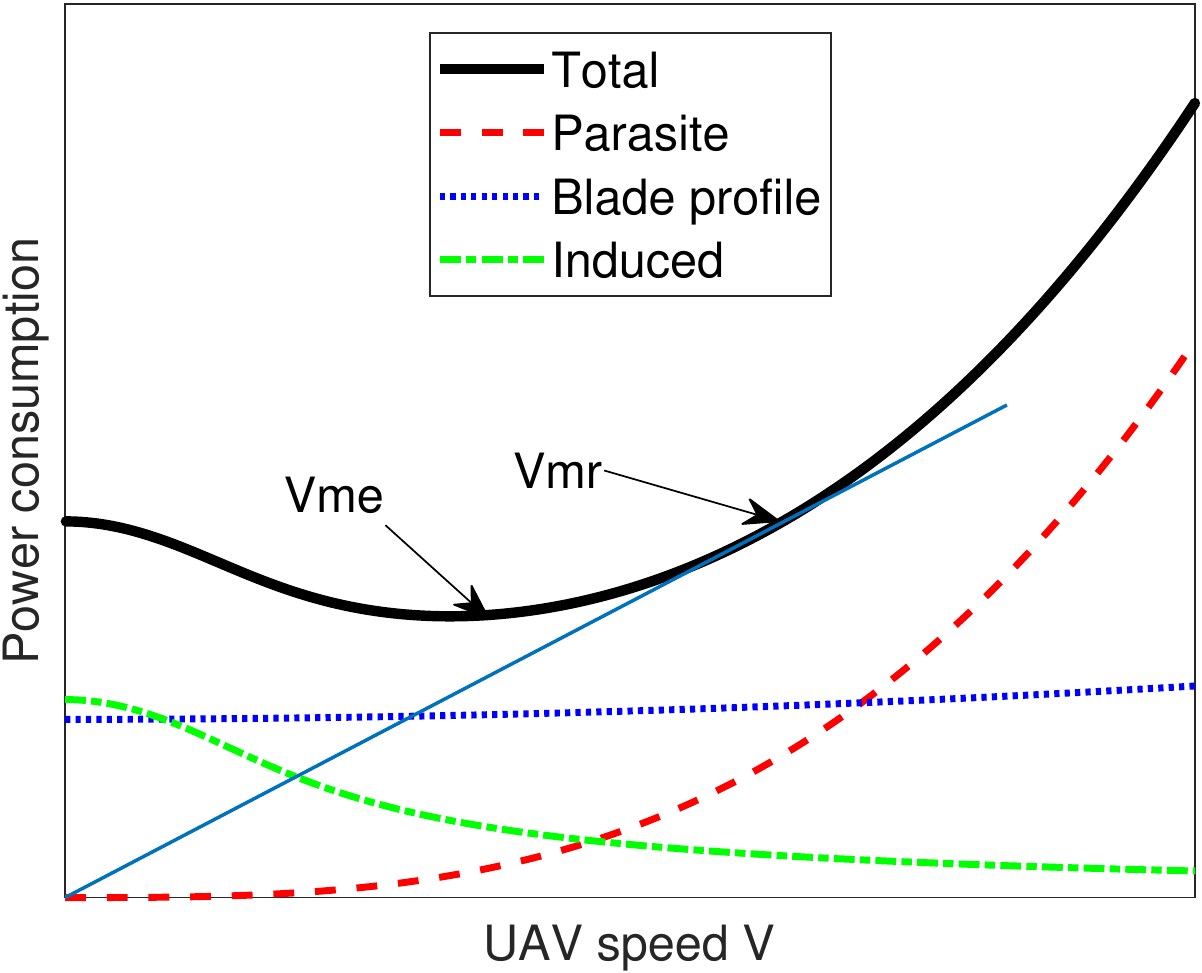}
  \caption{A typical plot of power consumption for rotary-wing UAV in straight-and-level flight \cite{8918497}.}
  \label{Fig:PowervsSpeedFixedWing2}
\end{figure}

The straight-and-level flight refers to the uniform rectilinear movement in the horizontal plane without UAV acceleration/deceleration. For a rotary-wing UAV in straight-and-level flight with speed $V$, it has been derived in \cite{8663615} that the power consumption can be expressed as
\begin{align}
P(V)=\underbrace{c_1(1+c_2V^2)}_{\text{blade profile}}+\underbrace{c_3\left(\sqrt{1 +\frac{V^4}{c_4^2}}-\frac{V^2}{c_4}\right)^{1/2}}_{\text{induced}}+\underbrace{c_5 V^3}_{\text{parasite}},
\label{eq:PVRotaryWing}
\end{align}
where $c_i, i=1,\cdots,5,$ are the modelling parameters that depend on the aircraft weight, air density, rotor disc area, as specified in \cite{8663615}. A typical plot of the power consumption curve (\ref{eq:PVRotaryWing}) is shown in Fig. \ref{Fig:PowervsSpeedFixedWing2}. Based on the figure and (\ref{eq:PVRotaryWing}), the following observations can be obtained \cite{8918497}.
\begin{itemize}
  \item The parasite power $c_5 V^3$ is the component required to overcome the parasite
friction drag due to the moving of the aircraft in the air, and the induced power $c_3\left(\sqrt{1 +\frac{V^4}{c_4^2}}-\frac{V^2}{c_4}\right)^{1/2}$ corresponds to that required to overcome the induced drag developed during the creation of the lift force to maintain the
aircraft airborne. Different from fixed-wing UAV, the blade profile power $c_1(1+c_2V^2)$ is a unique term for rotary-wing UAV, which is required to overcome the profile drag due to the rotation of blades.
  \item For the special case with speed $V=0$, the power consumption for rotary-wing UAV is given by a finite value $c_1+c_3$. This power consumption corresponds to the rotary-wing UAV hovering at fixed location.
  \item The parasite power exists only when the UAV has non-zero flying speed. Both the blade profile power and the parasite power are increasing with the aircraft speed $V$, while the induced power decreases as $V$ increases. The maximum-endurance speed $V_{\text{me}}$ is the optimal UAV speed that maximizes the UAV endurance for a given onboard energy and the maximum-range speed $V_{\text{mr}}$ is the optimal UAV speed that maximizes the total traveling distance with any given onboard energy \cite{8663615}.
\end{itemize}

\subsection{Experiment Setup}
\begin{figure}[!ht]
 \centering
  \includegraphics[width=7cm]{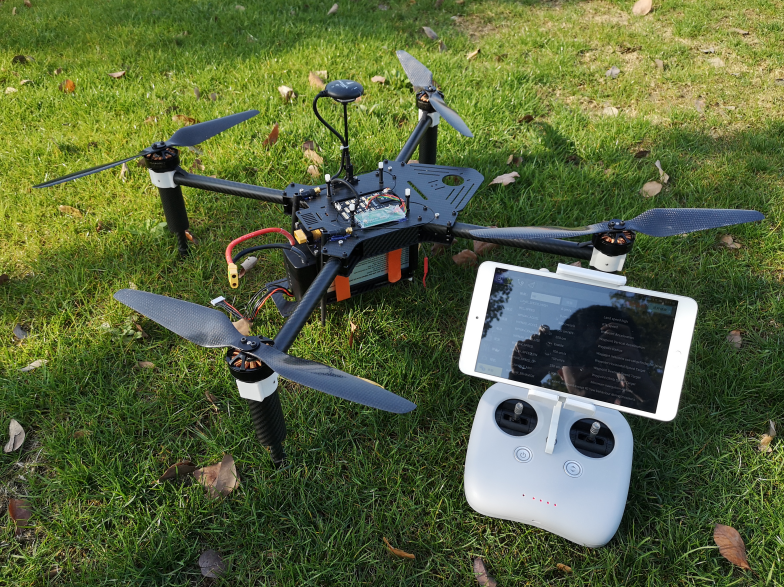}
  \caption{The UAV and its accessories used in the flight experiment.}
  \label{Fig:UAVphoto}
\end{figure}

To validate the above theoretical energy model, we have performed extensive flight experiments to actually collect the data for UAV power consumption. All experiments are performed on our customized quadrotor UAV. For all flight experiments, the instantaneous current flow and voltage of the UAV battery are recorded, together with the flying status, such as location, speed, acceleration, with a data collection frequency of 5Hz, i.e., one data measurement is obtained for every 0.2 seconds. Based on such data, the instantaneous total UAV power consumption is obtained, which includes both the flying power and the communication-related power. We firstly measured the UAV communication-related power consumption offline, with the UAV on the ground communicating with its control station, and found that the power level is around 0.8 W, which is much lower than the power consumption when flight in the air (several hundreds of watts). Therefore, the communication-related power is ignored and the recorded power consumption of the UAV battery treated as the flying power. The actual UAV and its accessories used in the flight experiments are shown in Fig. \ref{Fig:UAVphoto}, and the specifications of the UAV and the battery are listed in Table~\ref{Table:UAV} and
Table~\ref{Table:battery}, respectively.

The flight altitudes for all experiments are set to be 20 m. The first experiment focuses on the straight-and-level flight with an uniform speed. To validate the energy model (\ref{eq:PVRotaryWing}), we need to obtain different power consumption value at different flying speed. In each specified speed, the UAV is instructed to take off and fly following the pre-determined straight-and-level path for data collection. The experiment is repeated for each targeting speed. Ranging from 0 to 14 m/s in a step size of 1 m/s. Due to the site space restriction, the UAV needs to fly several round trips in order to collect enough data for each specified speed. Thus, with fixed data sampling rate, the number of required round trips vary with the UAV speed: the faster the speed is, the more flight round trips are needed.

\begin{table}\caption{The specifications of the UAV.} \label{Table:UAV}
\centering
\begin{tabular}{|c|c|}
\hline
\it Type & Quadrotor\\
 \hline
 \it Net weight without battery & 1.64 kg\\
\hline
 \it Height & 19.5 cm\\
\hline
 \it Rotor radius & 14 $\times$ 2.54 cm\\
\hline
 \it Rotor pitch & 5.5 $\times$ 2.54 cm\\
\hline
 \it Motor & KV460\\
\hline
\end{tabular}
\end{table}

\begin{table}\caption{The specifications of the battery.} \label{Table:battery}
\centering
\begin{tabular}{|c|c|}
 \hline
\it Weight & 1.36 kg\\
\hline
\it Capacity &  10000 mAh\\
 \hline
 \it Voltage &  3.5 $\times$ 6V $\sim$ 4.2 $\times$ 6V\\
\hline
 \it Dimensions & 17.5 $\times$ 6.5 $\times$ 5.5 cm\\
\hline
\end{tabular}
\end{table}

\subsection{Data Processing}\label{sec:DP}
\subsubsection{Data Preprocessing}
\begin{figure}[!ht]
 \centering
  \includegraphics[width=8cm]{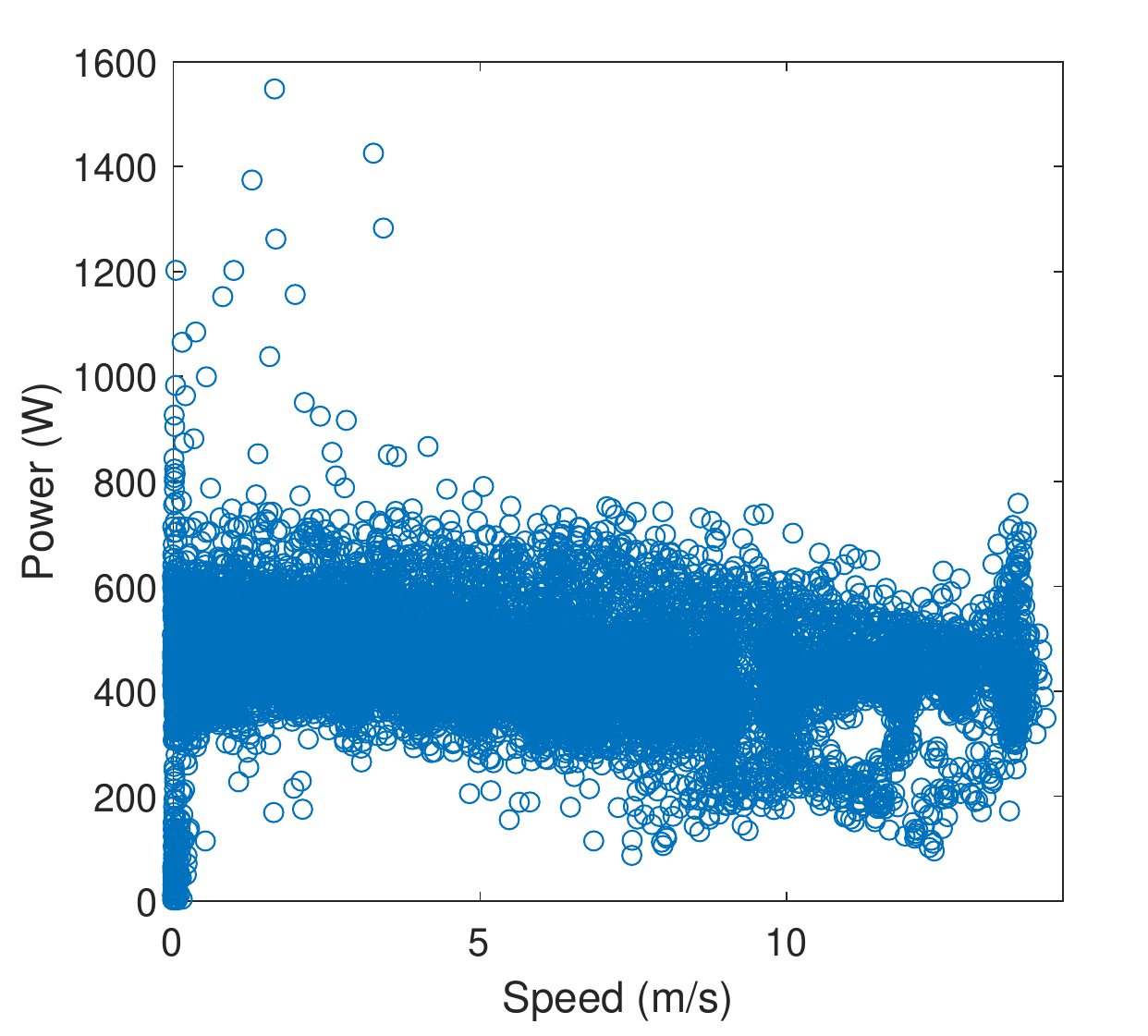}
  \caption{The scattering plot of the measured data in straight-and-level flight before data pre-processing.}
  \label{Fig:notProcessing}
\end{figure}

\begin{figure}[!ht]
 \centering
  \includegraphics[width=8cm]{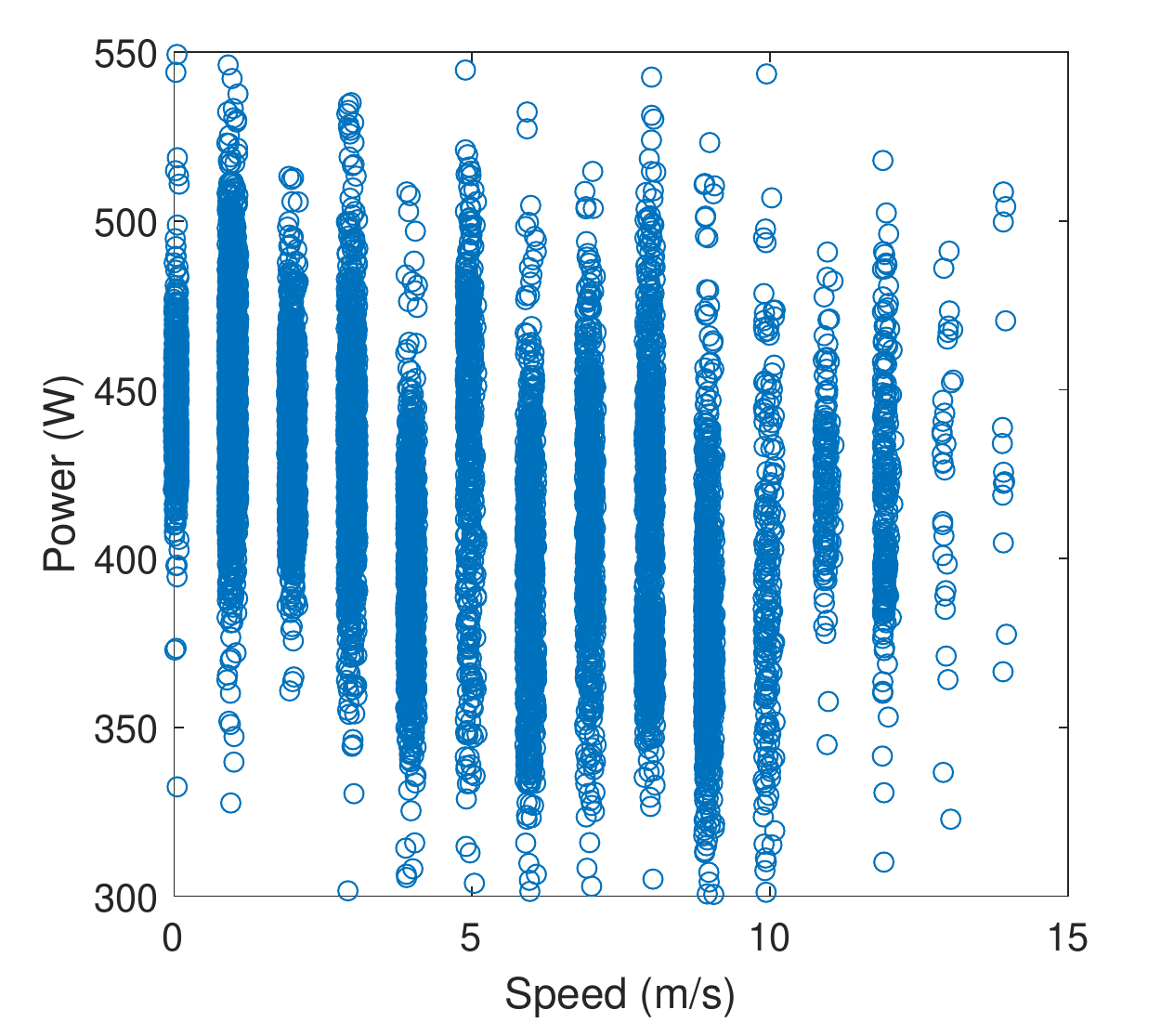}
  \caption{The scattering plot of the measured data in straight-and-level flight after pre-processing.}
  \label{Fig:afterProcessing}
\end{figure}

Note that the energy model in (\ref{eq:PVRotaryWing}) is derived for uniform flying as a function of its flying speed. Thus, among all the data measurements collected, we need to filter out those data when the UAV is in relatively steady flight. For the targeted straight-and-level flight, the raw measured data without any preprocessing is shown in Fig. \ref{Fig:notProcessing}. As the UAV needs to accelerate to the specified speed at the beginning of the experiment and decelerate to zero while it reaches the site edge and needs to turn around, there are some data points irrelevant to the specified speed. To this end, the data is pre-processed to exclude those points in the acceleration or deceleration phases. Specifically, the number of round trips in each experiment is recorded on the global position system (GPS) supported by google map. The data with respect to speed can be obtained from the google mobile service (GMS) of the GPS. Denote by $P_i$ and $P_{i+1}$ the measured power consumption at adjacent sampling intervals, and $V_i$ and $V_{i+1}$ the corresponding speed for $P_i,P_{i+1}$. Then the acceleration/decleration for the corresponding time interval can be approximated as
\begin{align}
a\approx(V_{i+1}-V_{i})f_s,
\label{eq:acc}
\end{align}
where $f_s=5$ Hz is the data sampling frequency. If $|a|>0.5~\text{m/s}^\text{2}$, we regard that the UAV is in acceleration/decleration phase; otherwise, it is in relative steady flight and the corresponding measurement data is retained for further processing. The measured data after such preprocessing is illustrated in Fig. \ref{Fig:afterProcessing}.
\subsubsection{Model-Based Curve Fitting}
Curve fitting aims to fit data to user-defined fitting functions. Therefore, we first use the experiment measurement data to fit our theoretical energy model in (\ref{eq:PVRotaryWing}). Denote the measured speed-power pair as $(V_i, P_i)$ for the $i$-th data point after preprocessing. With the least squares fitting, the goal is to find the modelling parameters $c_1,...c_5$ in (\ref{eq:PVRotaryWing}) by minimizing the mean square error, i.e.,
\begin{align}
\min_{c_j,j=1,\cdots,5} \sum_{i=1}^N[P_i-P(V_i)]^2,
\label{eq:ls}
\end{align}
where $N$ denotes the total number of valid data pairs after preprocessing, $P(V_i)$ represents power consumption obtained from the theoretical energy model given in (\ref{eq:PVRotaryWing}). Here, we use the built-in functionality of Matlab for curve fitting with the obtained data measurements.
\subsubsection{Model-Free Deep Neural Network Training}
\begin{figure}[!ht]
 \centering
  \includegraphics[width=8cm]{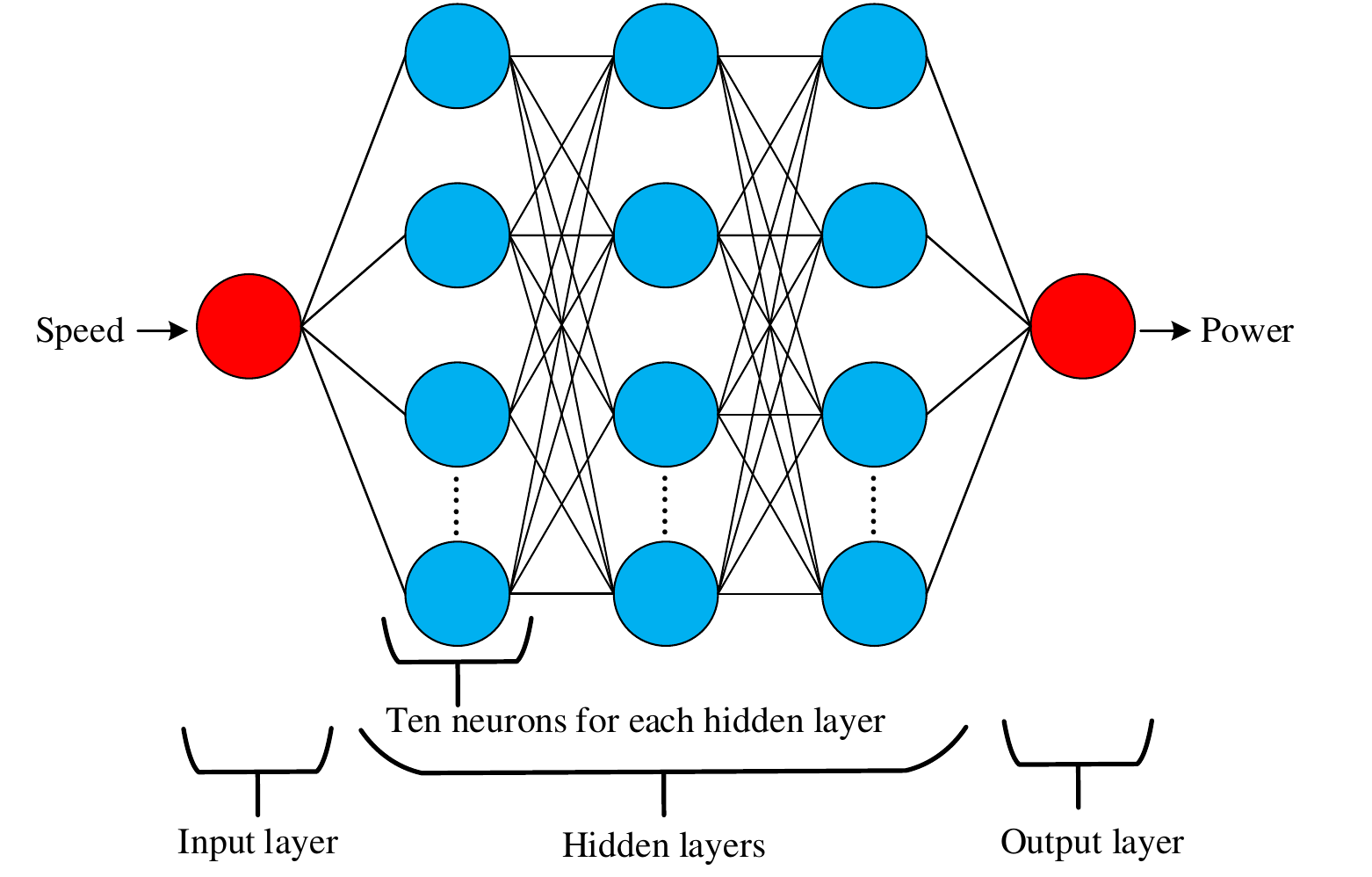}
  \caption{Deep neural network for model-free curve fitting of UAV energy consumption. The network takes the UAV flying speed as the input, and output the predicted power consumption.}
  \label{Fig:neural}
\end{figure}

To exclude the potential bias caused by the theoretical energy model, we also applied a model-free training with the obtained data based on a deep neural network. A neuron in a neural network is a mathematical function that collects and classifies information based on a specific architecture. The network bears a strong resemblance to statistical methods such as curve fitting and regression analysis. Thus, the neural networks are usually used for classification and regression tasks. Specifically, the neural network is regarded as a function approximator, and does not assume any prior model. By using a neural network with several hidden layers, the deep neural network can fit the nonlinear curves. The developed deep neural network architecture is shown in Fig. \ref{Fig:neural}. It consists of one input layer by taking the UAV speed as the input, three hidden layers with each hidden layer having ten neurons, and one output layer to predict the UAV power consumption. The $\mathrm{tanh}$ activation functions are used. The deep neural network is trained by using Tensorflow and Keras. In curve training, the learning rate is set to be 0.01. As a high-efficiency stochastic optimization algorithm, the Adam algorithm is used to train the network parameters to minimize the mean squared error loss.
\section{Experiment Results and Discussions}\label{sec:3}
\begin{figure}
\centering
\subfigure[GPS trajectory of straight-and-level flight.]{
\begin{minipage}[b]{0.5\textwidth}
\label{fig:straight1}
\centering
\includegraphics[width=6cm]{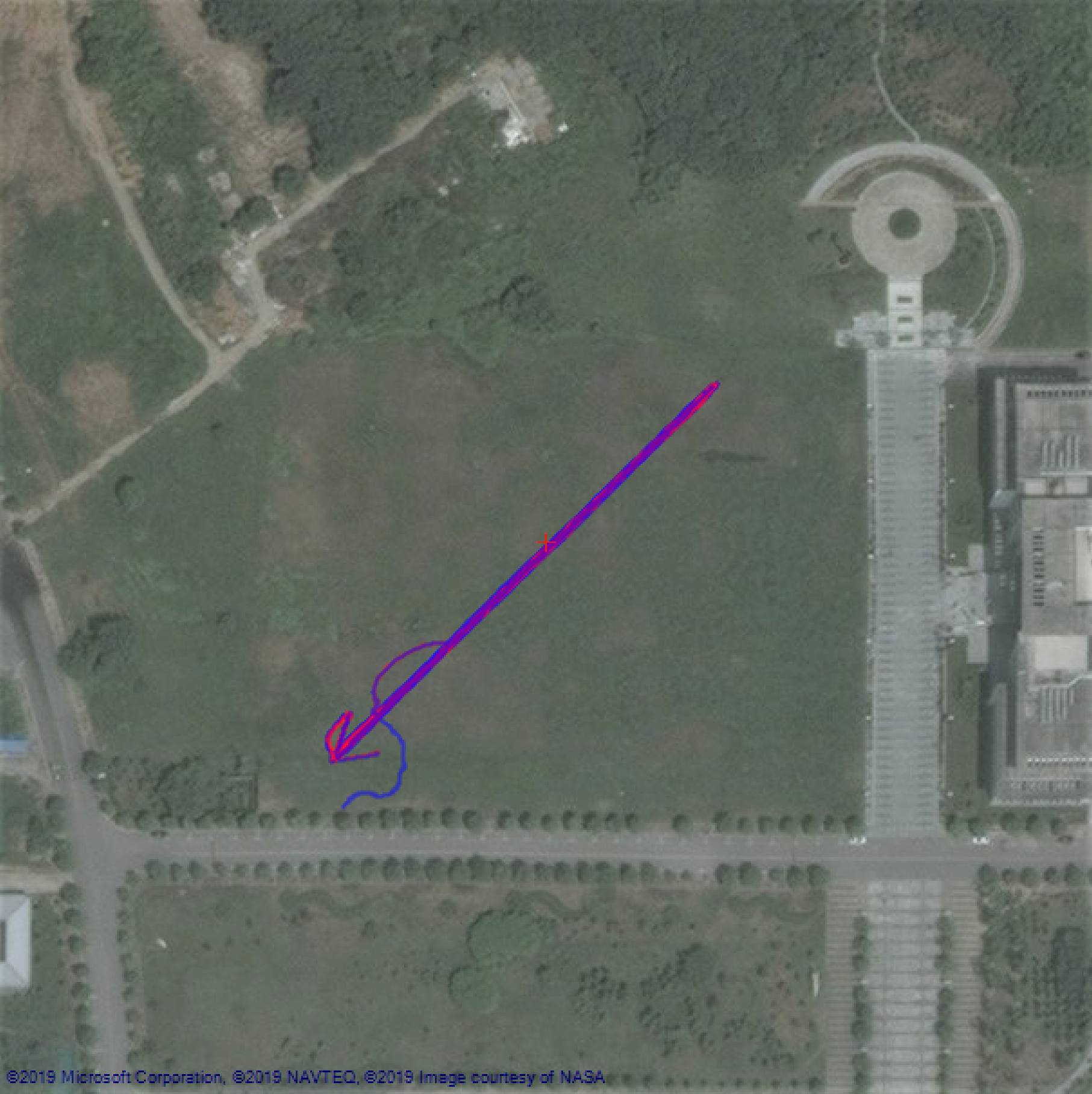}
\end{minipage}
}
\subfigure[UAV speed versus sample number.]{
\begin{minipage}[b]{0.5\textwidth}
\label{fig:straight2}
\centering
\includegraphics[width=8cm]{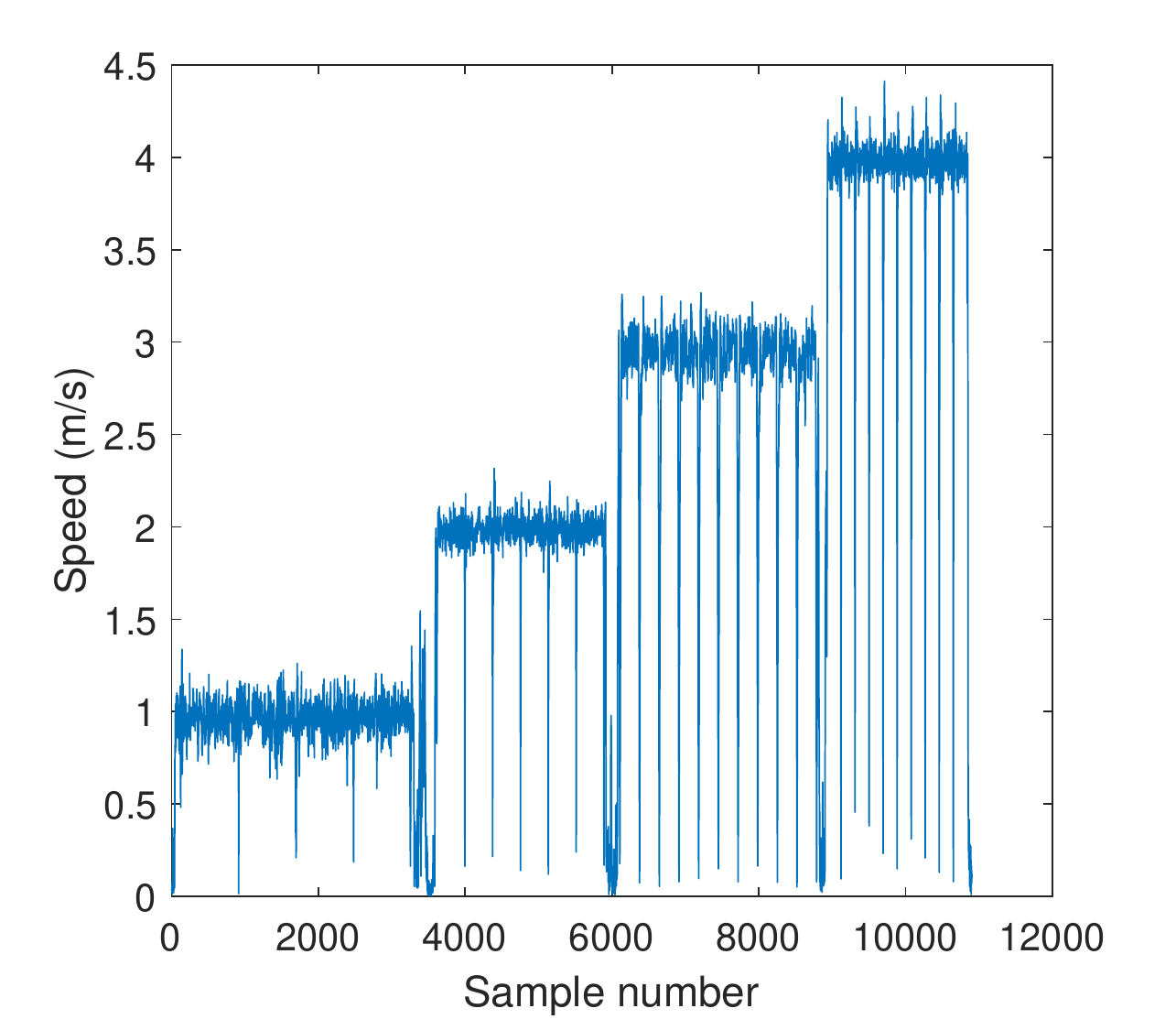}
\end{minipage}
}
\caption{Experiment 1: The straight-and-level flight experiment.}
\label{fig:1}
\end{figure}
\begin{figure}[!h]
 \centering
  \includegraphics[width=8cm]{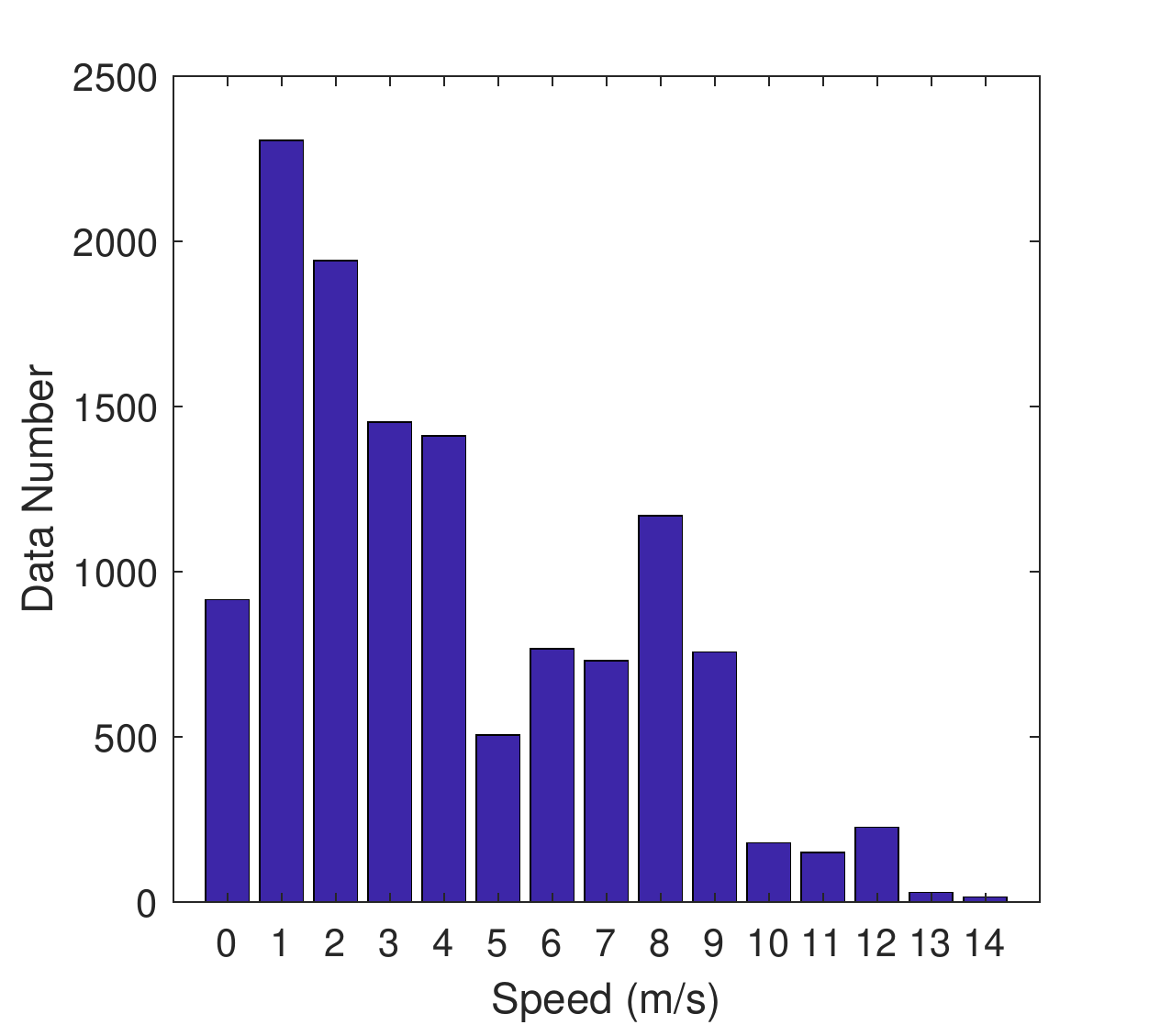}
  \caption{The number of valid data points after data preprocessing for different speed in straight-and-level flight.}
  \label{Fig:Amount_of_Data}
\end{figure}

We validate the theoretical energy model (\ref{eq:PVRotaryWing}) via the straight-and-level flight experiment, namely \textit{Experiment 1}. The GPS trajectory of one of the straight-and-level flights is shown in Fig. \ref{fig:straight1}. From the figure, we can see that the curved portion corresponds to the UAV taking off and landing phases, while the main long straight trajectories correspond to the straight-and-level flight. Fig. \ref{fig:straight2} shows one example of the recorded UAV speed versus the sample number for each specified speed. It is observed that for most of the time, the UAV speed fluctuates around the specified flying speed, while there are periodic abrupt speed changes to 0. These actually correspond to the direction reversal since the UAV has to accelerate/decelerate to reverse its flying direction when it reaches the boundary of the experiment site. Before performing curve fitting, we apply the data preprocessing as discussed in Section \ref{sec:DP} to retain those data in relatively steady flight, and the resulting number of data points for different speeds used for curving fitting is shown in Fig. \ref{Fig:Amount_of_Data}.

\begin{figure}[!ht]
 \centering
  \includegraphics[width=8cm]{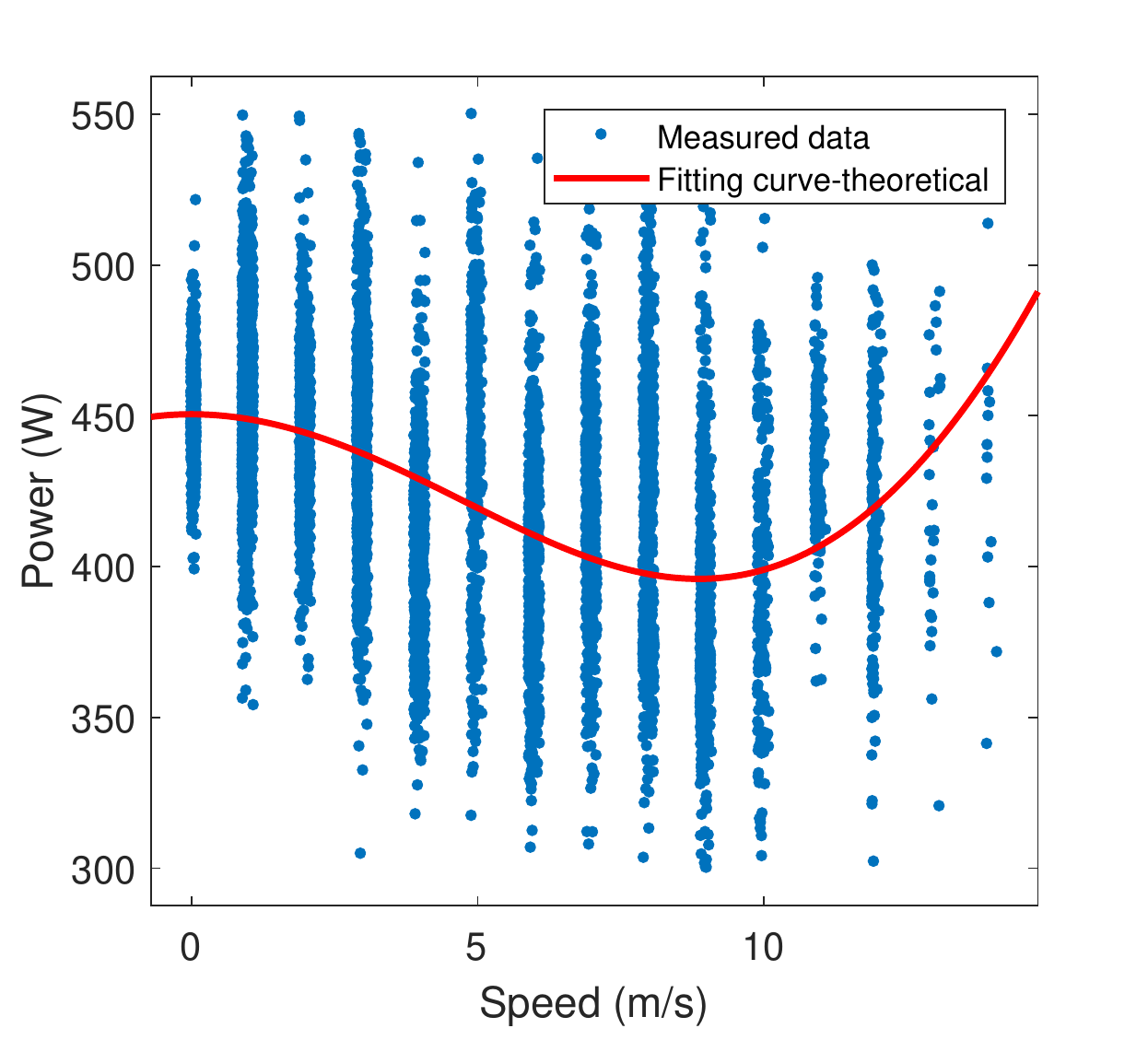}
  \caption{The obtained power versus speed relationship with curve fitting based on theoretical energy model (\ref{eq:PVRotaryWing}).}
  \label{Fig:fitness_custom}
\end{figure}

\begin{figure}[!ht]
 \centering
  \includegraphics[height=7.5cm,width=8cm]{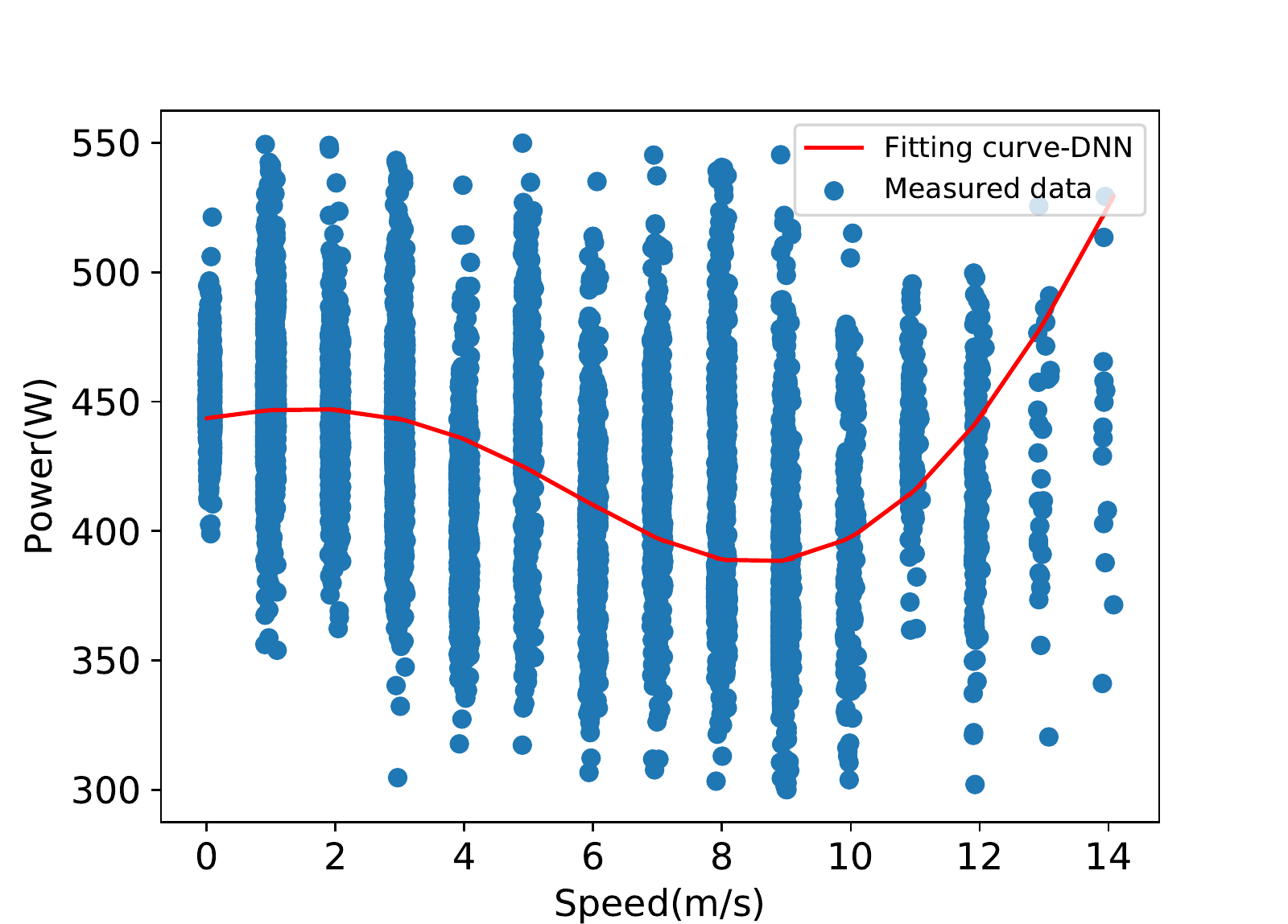}
  \caption{The obtained power versus speed relationship with curve fitting based on deep neural network training.}
  \label{Fig:Network1}
\end{figure}

\begin{figure}[!ht]
 \centering
  \includegraphics[width=8cm]{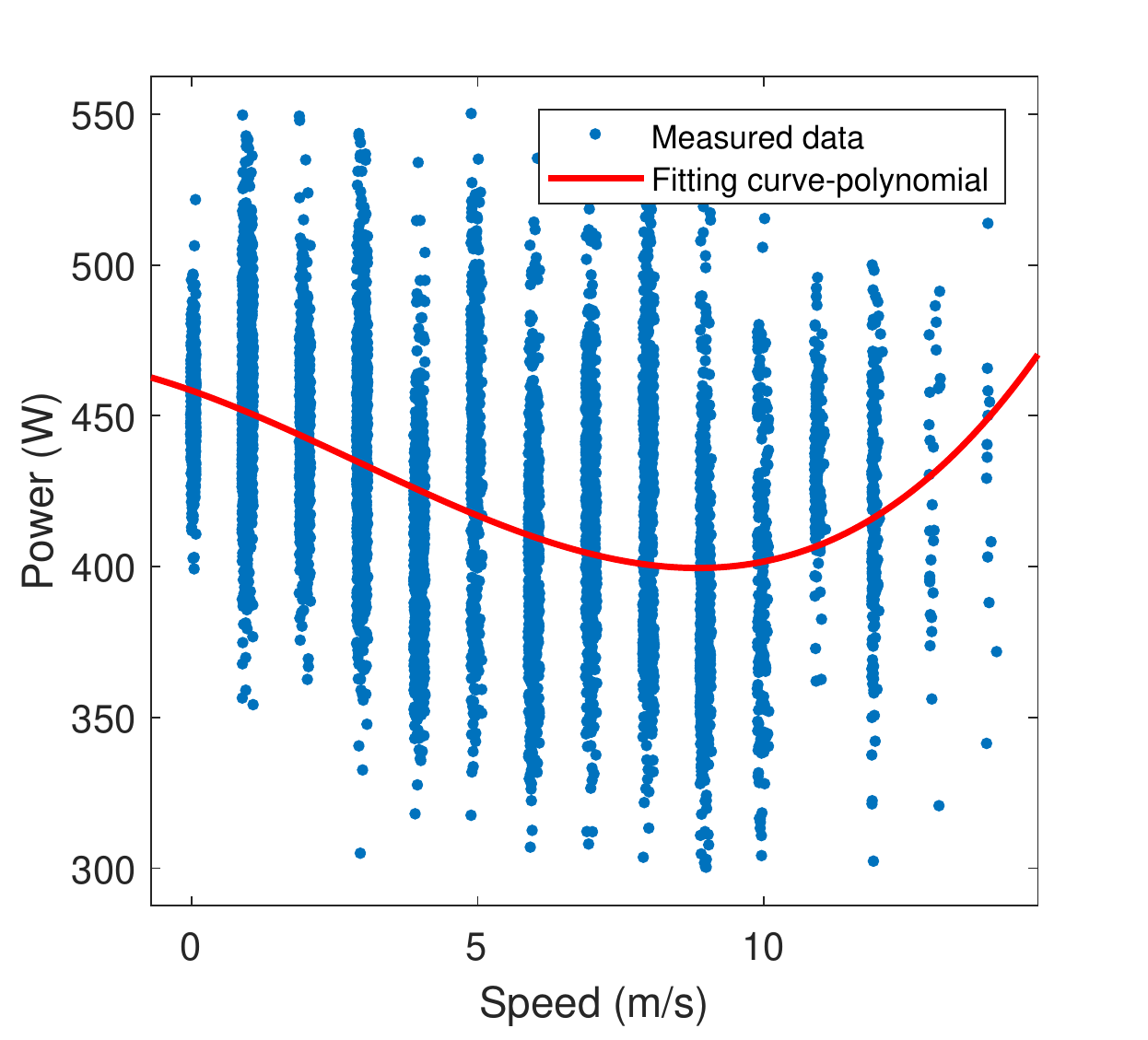}
  \caption{ The obtained power versus speed relationship with curve fitting based on polynomial model \cite{1302401}.}
  \label{Fig:fitness_polynomial}
\end{figure}

We apply the curve fitting based on the theoretical energy model (\ref{eq:PVRotaryWing}), and the result is shown in Fig. \ref{Fig:fitness_custom}. We can observe that the fitting curve based on theoretical energy model first increases then decreases, and then increases. With the flight speed increases, i.e., $V\geqslant10~ \text{m/s}$, the energy consumption increases significantly. Compared to Fig. \ref{Fig:PowervsSpeedFixedWing2}, we can find that the measured result matches quite well with the theoretical model curve.
To exclude the potential bias caused by the theoretical energy model, the energy consumption curve based on neural network training is used as a benchmark and the result is shown in Fig. \ref{Fig:Network1}. From the figure, we can find that the curve fitted by deep neural network also first increases then decreases, and then increases, which shows a good match with the curve based on the theoretical energy model. To further verify our theoretical energy model is better than the polynomial model in \cite{1302401}, the polynomial based curve fitting result is shown in Fig. \ref{Fig:fitness_polynomial}. The result shows that the power consumption curve first decreases and then increases, which cannot match well with both theoretical energy model based fitting curve and deep neural network based fitting curve, especially in low speed.
\section{Energy Model Generalization and Experimental Validation}
\subsection{Model Generalization for Arbitrary Flight}
Note that the energy model in (\ref{eq:PVRotaryWing}) is applicable only for straight-and-level flight with uniform speed. For more complicated scenario such as arbitrary 2-D flight, to our best knowledge, no rigorous theoretical
derivation has been reported for the closed-form energy model
as a function of its flying speed, direction, and acceleration. In this section, we extend (\ref{eq:PVRotaryWing}) to the arbitrary 2-D level flight by proposing the following heuristic energy model.

For arbitrary 2-D level flight with UAV trajectory $\mathbf q(t)\in \mathbb R^{2 \times 1}$, by following the similar decomposition of the UAV acceleration along the parallel and perpendicular directions of its flying velocity as in \cite{7888557}, the energy consumption of rotary wing UAV is presented at the top of the next page, where $v(t)= \dot q(t)$ is the UAV flying velocity, $a(t)=\ddot q(t)$ is the acceleration, $a^2_\bot(t)=\sqrt{\|\mathbf a(t)\|^2-\frac{(\mathbf a^T(t)\mathbf v(t))^2}{\|\mathbf v(t)\|^2}}$ is centrifugal acceleration , $g$ is the gravitational acceleration, and $\Delta_K=\frac{1}{2}m(\|\mathbf v(T)\|^2-\|\mathbf v(0)\|^2)$ is the change in kinetic energy.
\begin{figure*}
\begin{align}
E(\mathbf q(t))=\underbrace{
\int_0^T c_3\sqrt{1+\frac{a^2_\bot(t)}{g^2}}
\left(\sqrt{1+\frac{a^2_\bot(t)}{g^2}+\frac{\|\mathbf v(t)\|^4}{c^2_4}}-\frac{\|\mathbf v(t)\|^2}{c_4}\right)^{1/2}dt}_{\text{induced}}+\underbrace{\int_0^Tc_1\left(1+c_2\|\mathbf v(t)\|^2\right)dt}_{\text{blade profile}}\nonumber\\
+\underbrace{\int_0^Tc_5 \|\mathbf v(t)\|^3dt}_{\text{parasite}}+\Delta_K
\label{eq:arbitrary2D}
\end{align}
\hrulefill
\end{figure*}
From (\ref{eq:arbitrary2D}), we have the following observations:
\begin{itemize}
  \item The UAV's energy consumption for arbitrary 2-D level flight only depends on its speed $V(t)=\|v(t)\|$ and centrifugal acceleration $a_\bot(t)$. The centrifugal acceleration $a_\bot(t)$ is normal to the UAV velocity vector and causing heading change yet without altering the UAV's speed.
  \item The accumulative impact of tangential acceleration on UAV power consumption is reflected in the change of kinetic energy, which only depends on the initial and final speed,, rather than the intermediate speeds.
  \item One important special case of the arbitrary 2-D level flight is the circular level flight with uniform speed. Let $r$ denote the circle radius and $V$ the flying speed. Then the corresponding power consumption as a function of $r$ and $V$ can be obtained by letting $a_\bot=\frac{V^2}{r}, \|\mathbf{v}(t)\|=V, \forall t$.
\end{itemize}

\textit{Experiment 2} is set up to validate the heuristic energy model. For the special case of circular level flight with radius $r$ and speed $V$, where $V$ varies from 1 m/s to 6 m/s in step size of 1 m/s, and the flight radius $r$ changes from 10 m to 20 m in step size 10 m. For each specified speed and flight radius, the UAV is made to take off and fly in horizontal circular to obtain enough data points. Then, we repeat the experiment for the next specified speed and flight radius. In data preprocessing, the data points corresponding to taking off and landing phases are excluded, so as to only retain the data points corresponding to circular level flight.
\subsection{Experiment 2: The circular level flight}
\begin{figure}
\centering
\subfigure[GPS trajectory of circular level flight.]{
\begin{minipage}[b]{0.5\textwidth}
\label{Fig:circle1}
\centering
\includegraphics[width=6cm]{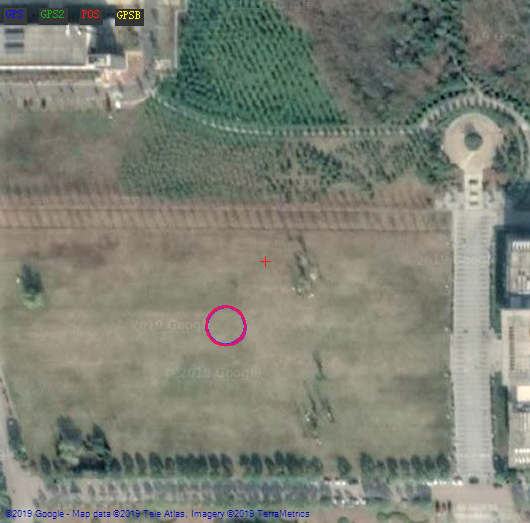}
\end{minipage}
}
\subfigure[UAV speed versus sample number for circular level flight]{
\begin{minipage}[b]{0.5\textwidth}
\label{Fig:circle2}
\centering
\includegraphics[width=8cm]{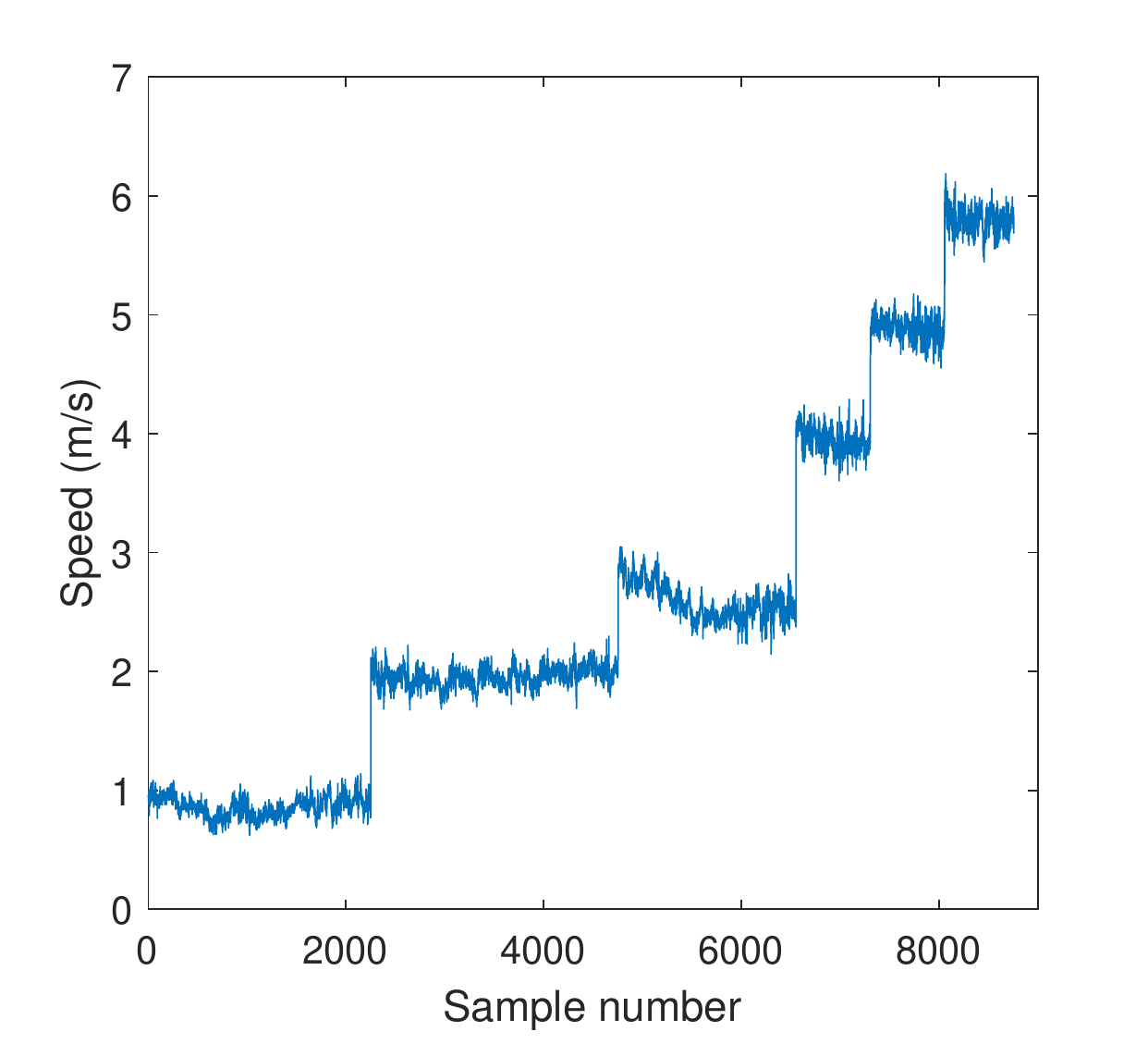}
\end{minipage}
}
\caption{Experiment 2: The circular level flight experiment.}
\label{Fig:circle}
\end{figure}
\begin{figure}[!ht]
 \centering
  \includegraphics[width=8cm]{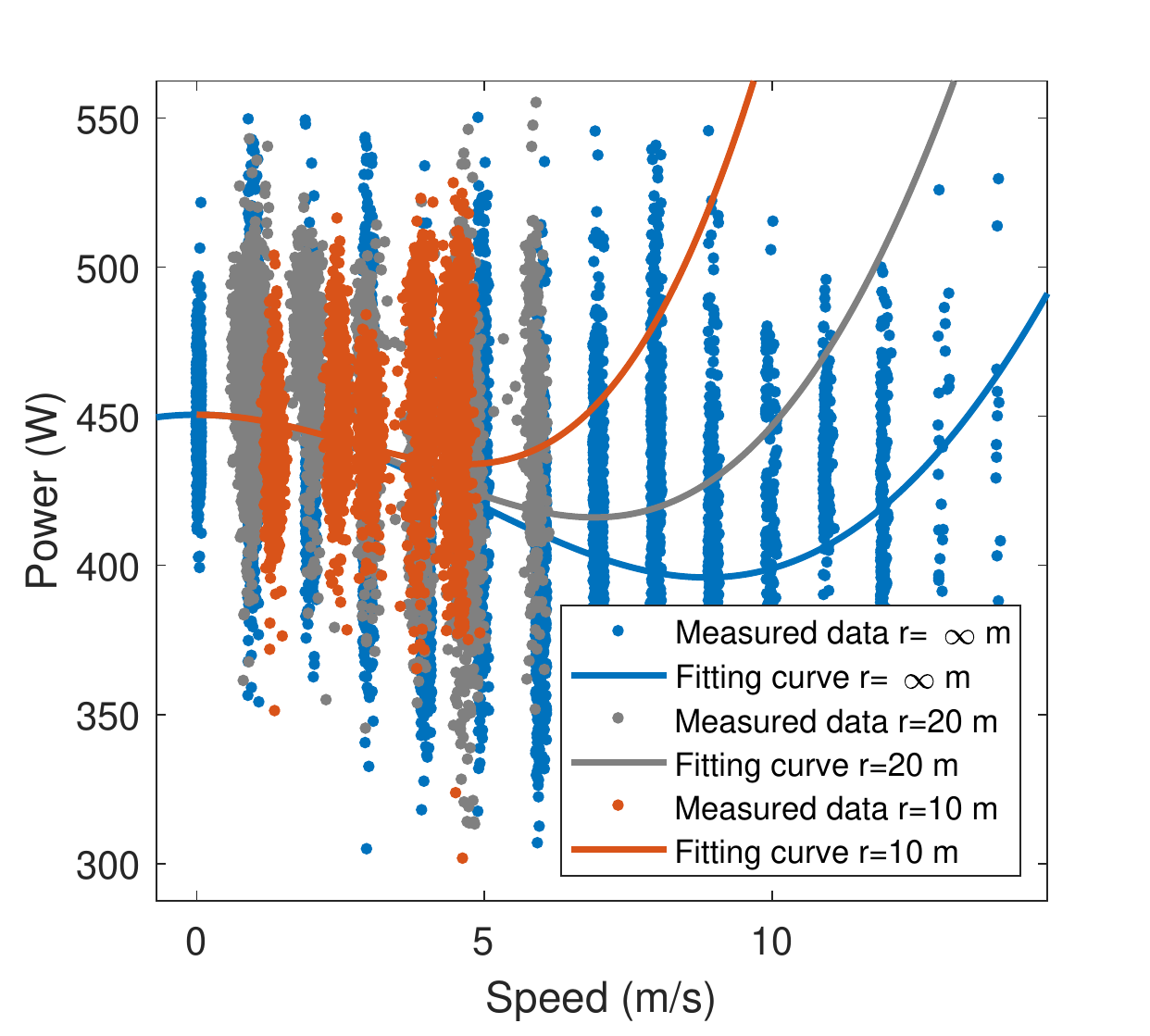}
  \caption{The fitting curves of power consumption in circular level flight.}
  \label{Fig:fitting_circular}
\end{figure}

The GPS trajectory of one of the circular level flight experiments is shown in Fig. \ref{Fig:circle1}, which is represented by the red circle. Fig. \ref{Fig:circle2} shows the recorded UAV flying speed versus the sample number. We can observe that the measurement speeds fluctuate around the specified speeds. The reason is that the measurement speeds are slightly affected by the complex environment factor such as the speed of wind. With the modelling parameters $c_1,...c_5$ obtained based on the curve fitting in straight-and-level flight discussed in Section \ref{sec:3}, resulting theoretical energy
model and measurement results for circular level flight are shown in Fig. \ref{Fig:fitting_circular} with different flight radiuses. The flight radius $r=\infty$ corresponds to special case of straight-and-level flight. From the figure, it is observed that the fitting results can well match the heuristic energy model of circular level flight. For fixed flight speed, we find that the power consumption increases as the circle radius decreases. This is because that the UAV has to consume more power to increase the centrifugal acceleration $a_\perp$ when the radius becomes smaller. This result can verify (\ref{eq:arbitrary2D}) that the centrifugal acceleration is inversely proportional to the flight radius. Compared with different curves, in low speed, i.e., $V\leq 4$ m/s, the flight radiuses have insignificant impact on power consumption, and with the flight speed increases, i.e., $V\geq 4$ m/s, the impact of flight radius on power consumption becomes more significant.

\section{Conclusions And Future Work}\label{sec:con}
In this paper, we have validated the recently derived theoretical energy model for rotary-wing UAVs via flight experiments. With the collected measurement data, we have used the model-based fitting and model-free deep neural network fitting to fit the curves. In model-based curve fitting, we have found that the fitted curves based on both the theoretical energy model and the model-free deep neural network match quite well with each other.  Furthermore, the theoretical energy model better characterizes  the UAV energy consumption than the polynomial model. In addition, we have proposed a heuristic energy model for arbitrary 2-D level flight, and verified the result for circular level flight.

In our future work, more experiments will be conducted to collect more data to further verify the accuracy of the theoretical energy model. To further generalize the energy model, we will study the energy model for arbitrary 3-D flight and design the corresponding flight experiments to validate its correctness. Furthermore, experiments by taking into account various other factors like wind speed, aircraft weight, altitude (air density), etc., will also be conducted for theoretical model verifications.




\ifCLASSOPTIONcaptionsoff
  \newpage
\fi



\bibliographystyle{IEEEtran}
\bibliography{IEEEabrv,sigproc} 
%

%








\end{document}